\documentclass[twocolumn,superscriptaddress,showpacs,floatfix,prb]{revtex4}
\usepackage{graphicx}
\begin{document}

\title{Impact of oxygen doping and oxidation state of iron on the electronic
  and magnetic properties of BaFeO$_{3-\delta}$
  }

\author{I.~V.~Maznichenko}
\email{igor.maznichenko@physik.uni-halle.de}
\affiliation{Institut f\"ur Physik, Martin-Luther-Universit\"at Halle-Wittenberg,
D-06099 Halle, Germany}

\author{S.~Ostanin}
\affiliation{Max-Planck-Institut f\"ur Mikrostrukturphysik, Weinberg 2, D-06120
Halle, Germany}

\author{L.~V.~Bekenov}
\affiliation{Institute of Metal Physics, Vernadsky Street 36, 03142 Kiev, Ukraine}

\author{V.~N.~Antonov}
\affiliation{Institute of Metal Physics, Vernadsky Street 36, 03142 Kiev, Ukraine}
\affiliation{Max-Planck-Institut f\"ur Mikrostrukturphysik, Weinberg 2, D-06120
Halle, Germany}

\author{I.~Mertig}
\affiliation{Institut f\"ur Physik, Martin-Luther-Universit\"at Halle-Wittenberg,
D-06099 Halle, Germany}
\affiliation{Max-Planck-Institut f\"ur Mikrostrukturphysik, Weinberg 2, D-06120
Halle, Germany}

\author{A.~Ernst}
\affiliation{Max-Planck-Institut f\"ur Mikrostrukturphysik, Weinberg 2, D-06120
Halle, Germany}

\date{\today}
\begin{abstract}
  We studied structural, electronic and magnetic properties of a cubic
  perovskite BaFeO$_{3-\delta}$ ($0 \le \delta \le 0.5$) within the
  density functional theory using a generalized gradient approximation
  and a GGA+U method.  According to our calculations, BaFeO$_3$ in its
  stoichiometric cubic structure should be half-metallic and strongly
  ferromagnetic, with extremely high Curie temperature ($T_C$) of 700
  -- 900~K. However, a such estimate of $T_C$ disagrees with all
  available experiments, which report that $T_C$ of the BaFeO$_3$ and
  undoped BaFeO$_{3-\delta}$ films varies between 111~K and 235~K or,
  alternatively, that no ferromagnetic order was detected there.
  Fitting the calculated x-ray magnetic circular dichroism spectra to
  the experimental features seen for BaFeO$_3$, we concluded that the
  presence of oxygen vacancies in our model enables a good agreement.
  Thus, the relatively low $T_C$ measured in BaFeO$_3$ can be
  explained by oxygen vacancies intrinsically presented in the
  material. Since iron species near the O vacancy change their
  oxidation state from $4+$ to $3+$, the interaction between Fe$^{4+}$
  and Fe$^{3+}$, which is antiferromagnetic, weakens the effective
  magnetic interaction in the system, which is predominantly
  ferromagnetic. With increasing $\delta$ in BaFeO$_{3-\delta}$, its
  $T_C$ decreases down to the critical value when the magnetic order
  becomes antiferromagnetic.  Our calculations of the electronic
  structure of BaFeO$_{3-\delta}$ illustrate how the ferromagnetism
  originates and also how one can keep this cubic perovskite robustly
  ferromagnetic far above the room temperature.
\end{abstract}
\maketitle

\section{Introduction}
\label{sec:Introduction}
A multiferroic tunnel junction, in which a robustly ferroelectric
barrier, such as PbTiO$_3$ or BaTiO$_3$, is grown epitaxially between
two ferromagnetic electrodes, allows to tune the spin transport by
switching the magnetization of an electrode. Additionally, the barrier
polarization reversal can change the interfacial magnetoelectric
coupling and, thus, the four-state tunneling resistance 
can be detected.\cite{Garcia26022010,valencia2011interface} As for the
upper electrode, the Fe and Co films can be easily grown on the
TiO$_2$ terminated polar perovskites,\cite{Meyerheim2011} while the
best bottom-electrode material, which can be used here as the
substrate, must satisfy the following conditions: (i) its structure
should be in a good lattice match with ferroelectrics and (ii) the
electrode material should be robustly ferromagnetic with relatively
high Curie temperature ($T_C$). A very good choice would be a magnetic
perovskite since a ferroelectric can be well grown on such
substrates.  In particular, perovskites {\it AB}O$_3$ with a
magnetic cation on site $B$ are well-known to be compatible with
ferroelectrics.\cite{MartinRamesh2012multiferroic} 
Recently, Pantel {\it et
  al.}\cite{pantel2012reversible} have demonstrated the four-state tunneling resistance 
effect using the 3.2-nm-thick perovskite barrier
PbZr$_{0.2}$Ti$_{0.8}$O$_3$~(001) grown on ferromagnetic
La(Sr)MnO$_3$ (LSMO).

One of the possible candidates for the substrate in a multiferroic tunnel junction seems to be
BaFeO$_3$ (BFO).  Although bulk BFO is hexagonal, the perovskite phase
can be stabilized in epitaxial films. X-ray diffraction studies show
that thin BFO films grown on SrTiO$_3$(001) adopt the cubic perovskite
structure with the lattice parameter of
3.97~\AA.\cite{chakraverty2013bafeo3,hayashi2011bafeo3} For
BaFeO$_{3-\delta}$ films, the larger lattice parameter of 4.07~\AA \
has been measured.\cite{ribeiro2012structural} 
Regarding the
observed magnetic properties of the cubic BFO, such as the Fe magnetic
moment, magnetic ordering and $T_C$, the reports seem contradictory.
Callender {\it et al.}\cite{callender2008ferromagnetism} reported that
a pseudocubic BFO is robustly ferromagnetic (FM) with
T$_C$=235~K. Instead, the oxygen deficient BaFeO$_{3-\delta}$ films
exhibit antiferromagnetism (AFM) at room temperature, while a weakly
ferromagnetic behavior extends above
390~K.\cite{ribeiro2012structural} For thick BFO films, Chakraverty
{\it et al.}\cite{chakraverty2013bafeo3} reported T$_C$=115~K and
a saturation magnetization of 3.2~$\mu_B$/f.u., which both remain stable
as a function of film thickness, with no signature of the spin spiral
structure up to 300~nm.

In the case of completely oxidized stoichiometric cubic BFO, the
oxidation state of iron is 4+ that leads to the twofold orbitally
degenerated configuration $t_{2g}^3 e_g^1$. In polyatomic molecules
the half occupied $e_g$ level have to split according to
Jahn and Teller.\cite{JahnTeller1937} In solids the effect must be cooperative.
However, for $A$FeO$_3$ ($A$
= Ca, Sr, Ba) this behavior, known as the cooperative Jahn-Teller
distortion, was not reported so far. 
Instead, it occurs in
antiferromagnetic LaMnO$_3$ where up to 750~K the $e_g$ splitting opens the band
gap.\cite{Rodriguez1998,Huang1998,Hemberger2002}
On the other hand, the Fe oxidation state in
oxides is so unstable that the presence of the degenerated $e_g$ level
is not required. For instance, in metallic CaFeO$_3$ the charge
disproportionaly is observed: 2Fe$^{4+}$ $\rightarrow$
Fe$^{(4-\tau)+}$ + Fe$^{(4+\tau)+}$, where $0 < \tau \leq
1$. \cite{hayashi2013field} Thus, the issue of mixed valency might be
very important in the case of BFO.
 
Several theoretical studies of magnetic BFO have been recently made
from first principles.  For the FM order and experimental lattice
parameter of the cubic BFO, the calculated magnetic moment varies
between 3.0~$\mu_{B}$ and 3.4~$\mu_{B}$ per f.u.,
\cite{fuks2013ab,li2013electronic} depending on the model
approximation used to treat the exchange-correlation effects.
According to the {\it ab initio} investigation by Ribeiro {\it et
  al.}\cite{ribeiro2013self} the FM solution in the stoichiometric BFO
is energetically favorable against the possible AFM configurations
while the FM moment is about 3.6~$\mu_{B}$.  A helical magnetic
ordering in BFO, which is usually discussed in the context of
SrFeO$_3$ and CaFeO$_3$,\cite{mostovoy2005helicoidal} has been
investigated from first principles by Zhi Li {\it et
  al.}\cite{li2012first,li2012pressure} 
The authors anticipate that
the G-type helical solution can be fixed, keeping the reduced volume
and, simultaneously, using the Hubbard parametrization which treats
electronic correlations beyond the standard of the local density
approximation. The theoretical estimations of $T_C$ were not performed
so far.
   
As it was already mentioned above, the Fe atoms in the BFO can have
various oxidation states. The origin of this distinction in real
materials can be related for instance with a particular sample
preparation procedure, that can explain the large discrepancy in
results of various experiments. In this work, we investigate from
first-principles the impact of Fe oxidation states on structural,
magnetic, optical, and electronic properties of chemically perfect BFO and
BaFeO$_{3-\delta}$ ($\delta > 0$) with vacancies. The paper is organized as follows. In Section~\ref{sec2} we
provide the essential details of our methods and calculations.  In
Section~\ref{sec3}, we present the results and discuss the structural,
electronic, optical, and magnetic properties of the BFO.  Finally, our
summary and conclusions are drawn.

\section{Details of calculations}
\label{sec2}
Our calculations of the BFO, undoped BaFeO$_{3-\delta}$, and nominally
overdoped BFO were performed using a first-principles Green function
method~\cite{Luders2001} based the multiple-scattering within the
density functional theory in a generalized gradient approximation (GGA)
to the exchange-correlation potential.~ \cite{Perdew1996} We used a
full charge density approximation to take into account possible
non-sphericity of the crystal potential and the charge density.

The Fe oxidation state was modeled using a self-interaction correction
(SIC) method~\cite{Perdew1981} as it is implemented within the
multiple-scattering theory.~\cite{Lueders2005} In this case, an
oxidation state can be defined by the number of valence electrons as
\begin{equation}
  N_{val} = Z - (N_{core} + N_{SIC}),
\end{equation}
where $Z$ is the total number of electrons of an isolated atom,
$N_{core}$ is the number of electrons occupying the deep energy levels
and treated as the core electrons, $N_{SIC}$ is the number of
SIC corrected 3$d$ electrons.
 
Since the SIC approach is designed only for ground state
properties,\cite{Perdew1981} we used a GGA+U
method~\cite{Anisimov1991} to take into account electronic correlation
effects in calculations of exchange parameters. The later were
estimated within the magnetic force theorem.~\cite{Liechtenstein1987}
The
critical temperature and the magnetic ordering were obtained within
the Heisenberg model
\begin{equation}
  H=-\sum\limits_{i,j} J_{ij} \, \mathbf{e}_i \cdot \mathbf{e}_j,
\end{equation}
where $i$ and $j$ label the magnetic atoms, and $\mathbf{e}_i$ is a
unit vector in the direction of the magnetic moment of the {\it i}\,th
atom.  The critical temperatures were estimated using both the mean
field approximation (MFA) and random phase approximation
(RPA).~\cite{Tyablikov1967} Both approaches were successfully applied to
many magnetically ordered oxides including magnetic
perovskites.~\cite{Fischer2009,Etz2012}

Disorder effects in our study were simulated within the
coherent-potential approximation (CPA).~\cite{Gyorffy1972,Oguchi1983,Gyorffy1985}
The intrinsic oxygen vacancies in BaFeO$_{3-\delta}$ were modeled
within the same CPA scheme, which implies randomly distributed
vacancies over all oxygen sites. Regarding the Ba and Fe vacancies, in
the literature, such cation-deficient perovskites are formally denoted
as BaFeO$_{3+\delta}$, although the bulk perovskite structures are
unable to accommodate interstitial oxygen.

The calculations of the XAS and XMCD 
were performed using the spin-polarized fully
relativistic linear-muffin-tin-orbital (LMTO) method
\cite{And75,NKA+83} for the experimentally observed lattice constants.
The details are 
described in our
previous papers.~\cite{ADK+05,AJY+06,AYJ10}
\section{Results}
\label{sec3}
\subsection{Crystalline structure of  BaFeO$_3$}
\begin{figure}
  \centering
  \includegraphics[width = .5\columnwidth]{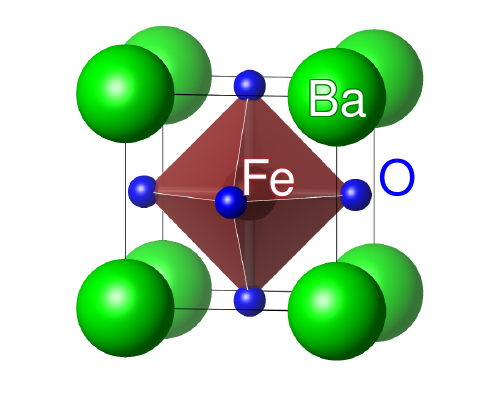}
  \caption{(Color) The crystalline structure of the cubic BaFeO$_3$.}
  \label{fig1}
\end{figure}

The ground-state phase of BFO is hexagonal with the space group
symmetry $P12_1/c1$.\cite{Zou1993} It contains six formula units
(f.u.) and two types of environmentally different Fe. We calculated
the structural properties using both the LDA \cite{Perdew1992} and GGA \cite{Perdew1996} to the
exchange-correlation potential. The GGA (LDA) yields the equilibrium
volume of 378.66~\AA$^3$ (352.84~\AA$^3$), which is in a good
agreement with the experimental value of 388.42~\AA$^3$. For this
volume and chemically perfect BFO, we found that the AFM configuration
is marginally preferable by 9~meV/f.u., as compared to the FM
solution, while the magnetic moments of each type of Fe differ
significantly: 3.34~${\mu}_B$ and 2.38~${\mu}_B$. The use of the
electronic-correlation parameter U=2~eV applied on iron species
increases the Fe magnetic moments by 12\%, whereas the energy
difference $\Delta E = E_{FM} - E_{AFM}$ increases up to 28~meV/f.u.

Concerning the cubic BFO phase, the LDA equilibrium lattice parameter
of 3.86~\AA \ is significantly underestimated, as compared to the
experimental value of 3.97~\AA, while the use of GGA results in
$a = 3.963$~\AA. For the latter we found that the FM ordering is
strongly favorable by 110~meV/f.u. with respect to the G-type AFM
configuration. Thus, we suggest that the BFO phase transformation from
its hexagonal AFM to cubic FM structure goes with the volume reduction
of 1.1\% per f.u.  and the corresponding energy barrier is 0.28~eV/f.u. In BFO films, the cubic FM phase can be stabilized at room
temperature by anharmonic contributions to the free energy and/or the
effects of epitaxial strain.  

Thus, the crystalline structure of the hexagonal and cubic BFO can be
adequately described using a conventional GGA functional. Further,
we focus our study on electronic, optic, and magnetic properties of
the cubic BFO phase (see Fig.\ref{fig1}). 

\subsection{Electronic and magnetic structure of  BaFeO$_3$ }
\subsubsection{GGA approach}
\begin{figure}
  \centering
  \includegraphics[width = .95\columnwidth]{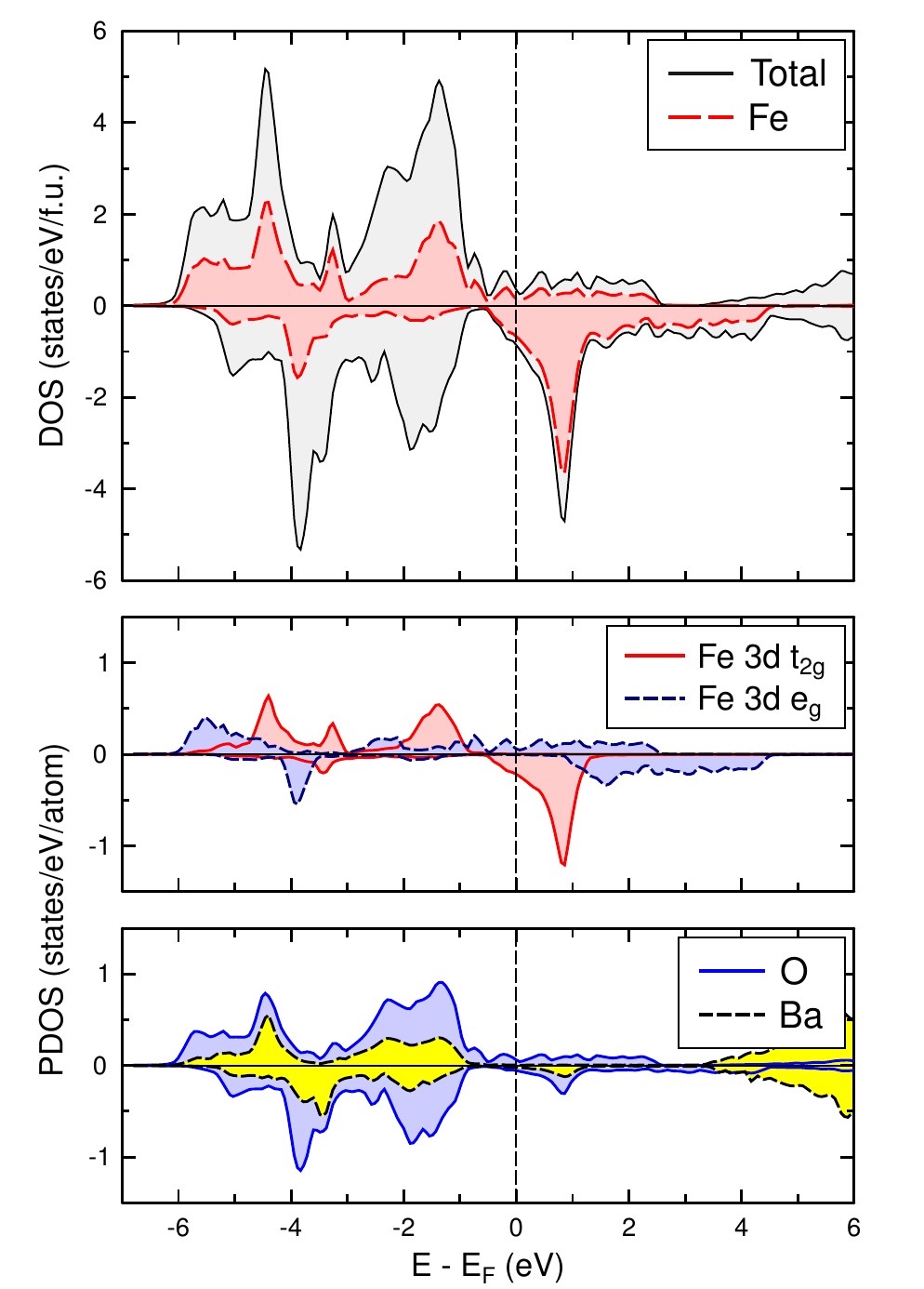}
  \caption{(Color) The total and site-projected DOS of cubic
    BFO calculated within the GGA. The positive (negative) DOS values denote
    the majority (minority) spin states.}
  \label{fig2}
\end{figure}

\begin{figure}
  \centering
  \includegraphics[width = .95\columnwidth]{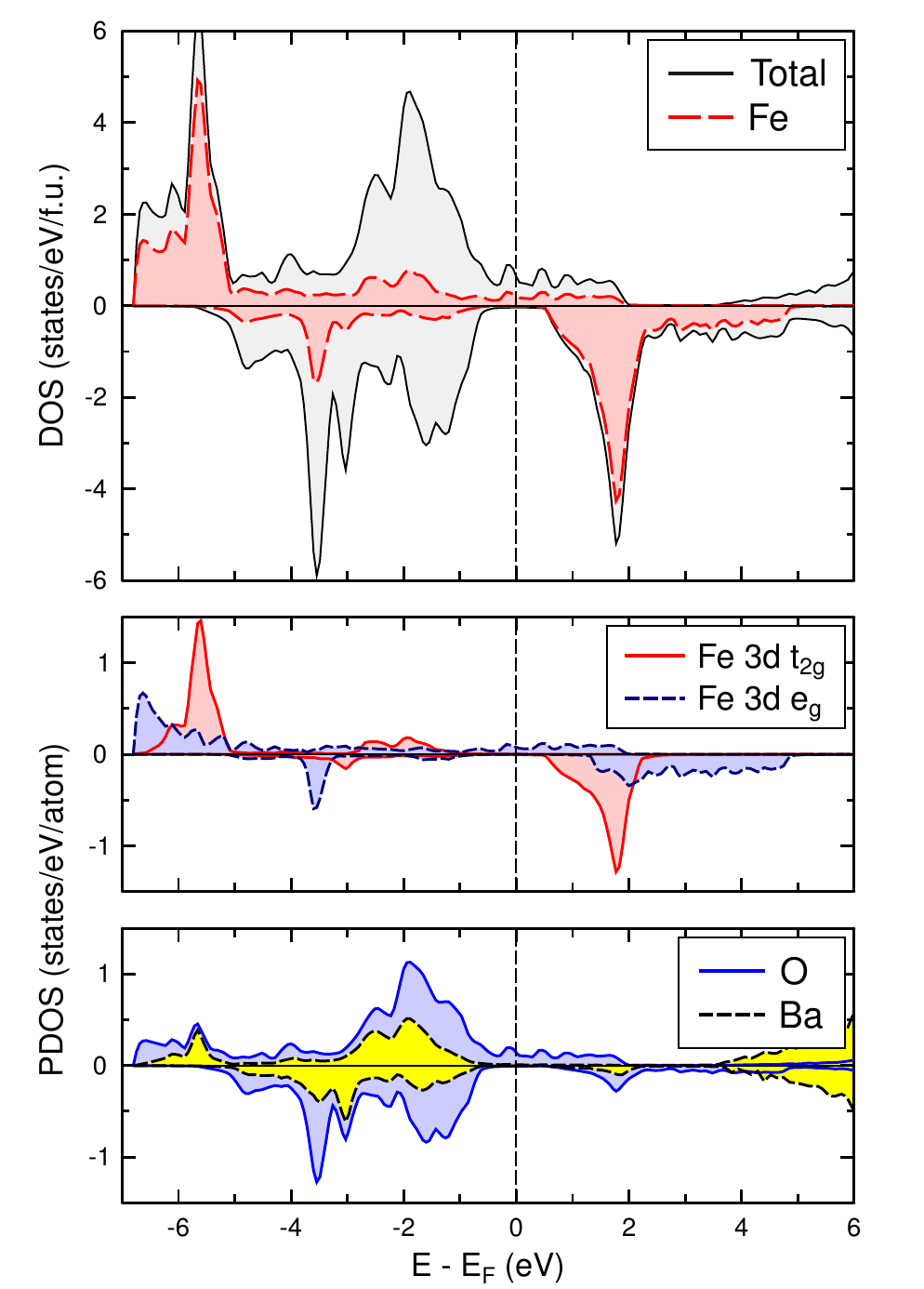}
  \caption{(Color) The total and site-projected DOS of cubic
    BFO calculated within the GGA+U method with $U_{eff}$=3~eV.}
  \label{fig3}
\end{figure}

Since we found the GGA to be an appropriate functional in describing
the crystalline structure of the cubic BFO, we applied the method for
further study of its electronic and magnetic properties. 

Within the GGA, the spin-polarized density of states (DOS) of BFO
exhibits a metallic behavior in the two spin channels, as it is shown
in Fig.\ref{fig2}. This agrees with the experimental findings of
Callender {\it et al.},\cite{callender2008ferromagnetism} who
observed the ferromagnetic ordering and conductance in BFO films,
annealed in the oxygen atmosphere. The iron atoms appear to have the
oxidation state of $2+$. Therewith, $t_{2g}$ and $e_g$ states are
almost occupied in the majority spin channel (4.5 electrons) and only
partially occupied (1.5 electrons) in the minority spin channel.  The
$3d$ electrons of iron experience a strong hybridization with the $p$
states of oxygen and are widely spread over the whole valence band
region.  The total magnetic moment, calculated within the GGA, is 
3.55~$\mu_{B}$ per f.u., which includes the ferromagnetically
induced O moments of about 0.15~$\mu_{B}$.

The exchange interaction between Fe atoms is large and positive within
the GGA: 20.5~meV and 2~meV between the nearest and the next nearest
neighbors, respectively. The estimated Curie temperature is
about 1000~K in the mean field approximation (700~K in the random
phase approximation). It has to be mentioned here that the RPA
provides usually a reasonable critical temperature, while the MFA
represents average exchange interaction and overestimates $T_C$ by
30 -- 40\%.\cite{Pajda2001}

Thus, the Curie temperature, calculated in the cubic BFO using
the GGA method, is several times larger then the experimental values
(100 -- 400~K). The reason is that the system is
metallic within this approach and, therefore, the double exchange
interaction via the oxygens and the band ferromagnetism dominate over
the other exchange mechanisms, which can reduce the magnetic
interaction or change its sign. 
Therefore, the electronic correlations,
which can be substantial in oxides, must be taken into account using a
more appropriate method.
\subsubsection{GGA+U approach}
At the next step, we chose the GGA+U approximation, which
provides a more adequate description of oxide materials then the
conventional LDA or GGA approaches.\cite{Anisimov1997} Since we could not
estimate the value of $U$ from first principles, we used it as a
parameter tracing the change of the electronic and magnetic properties
with the value of $U$. In our simulations we applied the Hubbard $U$
corrections on Fe $3d$ states and varied the value of $U_{eff}=U-J$
between 0 (here, GGA) and 9~eV. As expected, the occupied (unoccupied)
$3d$ Fe states are shifted down (up) in energy on the value of
$U_{eff}/2$ that opens the gap in the minority spin channel for
$U_{eff}\ge 2$~eV (see the DOS in Fig.\ref{fig3} with
$U_{eff}=3$~eV). The size of the spin band gap depends strongly on the
$U_{eff}$ value. 

\begin{figure}
  \centering
  \includegraphics[width = .95\columnwidth]{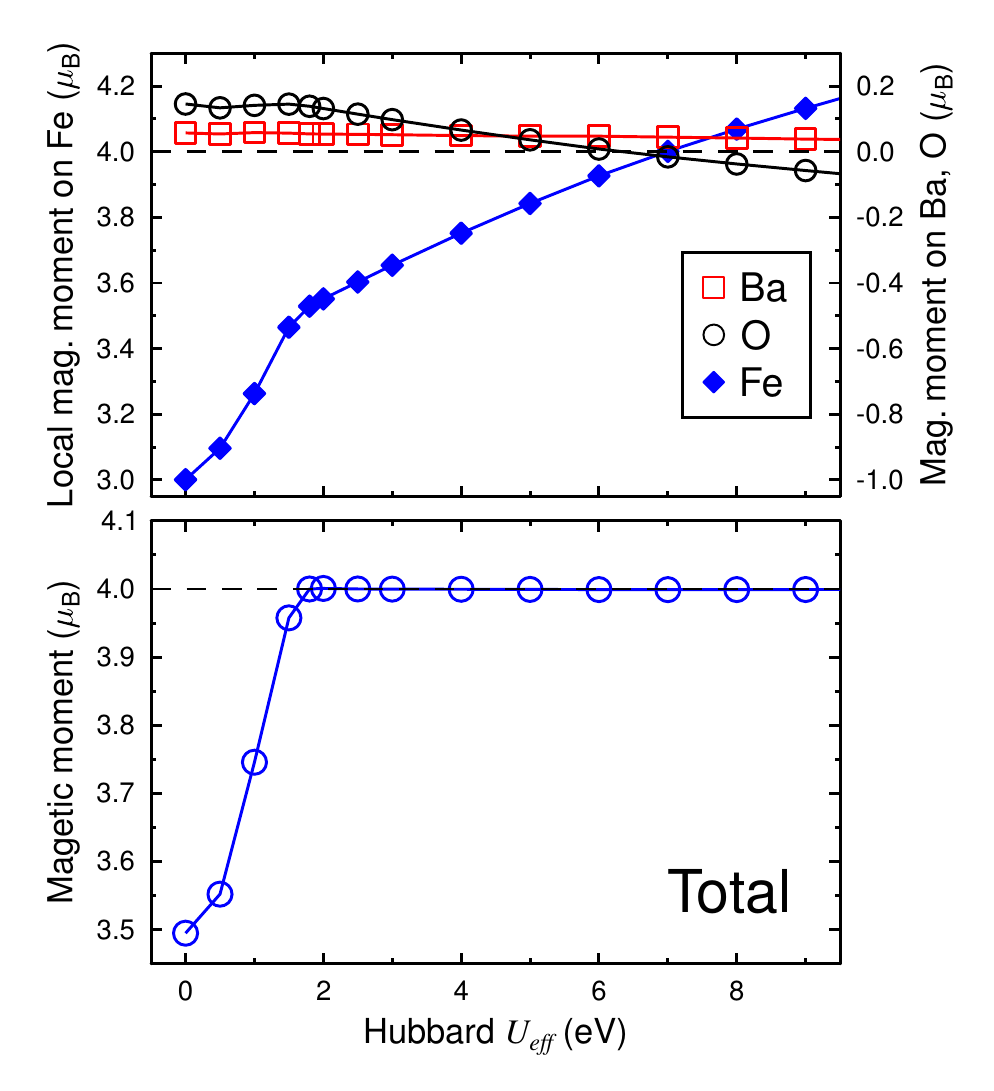}
  \caption{(Color) The total and site-projected magnetic
    moments of cubic BFO calculated as a function of $U$. Here,
    $U_{eff}=0$ means the use of GGA.}
  \label{fig4}
\end{figure}

In Fig.\ref{fig4}, we show the total and all
site-projected magnetic moments, which are plotted as a function of
$0\le U_{eff} \le 9$. The local magnetic moment of iron increases from
3~$\mu_B$ to 4.18~$\mu_B$ within the given range of $U_{eff}$ values,
while the induced magnetic moment of oxygen decreases from
0.18~$\mu_B$ to -0.08~$\mu_B$ systematically with the increase of
$U_{eff}$.  The magnetic moment induced on Ba is
$\approx 0.05~\mu_{B}$ and its small value changes marginally on the
model variations of $U$.  For $U_{eff}\ge 2$~eV the total moment is
integer (4~$\mu_{B}$/f.u), which reflects the half-metallicity of the
BFO. This is larger then the experimental values of 3.2 -- 3.5~$\mu_{B}$/f.u., reported previously in
Refs.~\onlinecite{hayashi2011bafeo3, chakraverty2013bafeo3}. 

\begin{figure}
  \centering
  \includegraphics[width = .95\columnwidth]{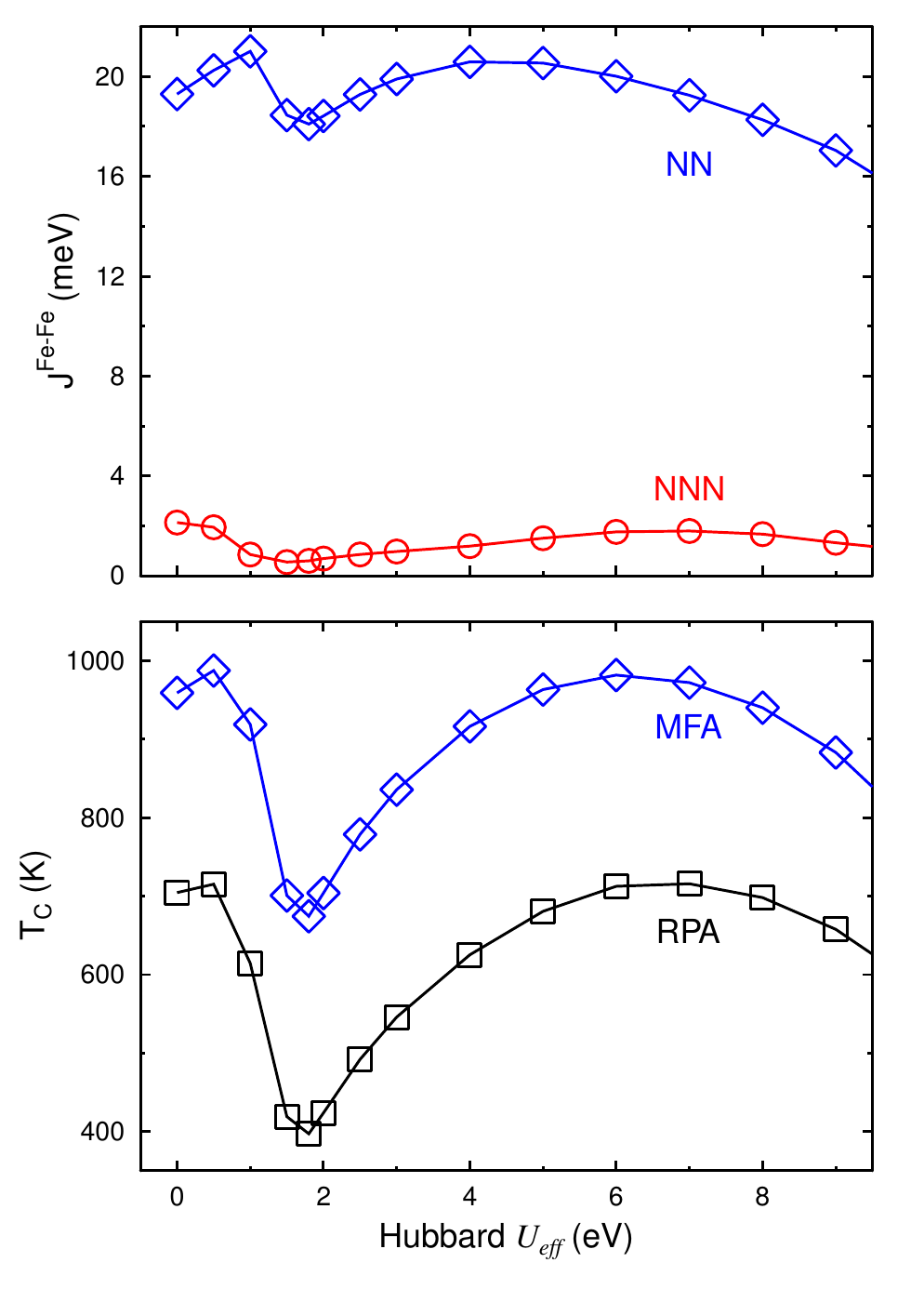}
  \caption{(Color) Exchange interactions (upper panel) and the
    Curie temperature (lower panel) of the cubic BFO calculated as a
    function of the Hubbard parameter $U_{eff}$ applied on the Fe $3d$
    orbitals.}
  \label{fig5}
\end{figure}

As next, we calculated the exchange interaction as a function of
$U_{eff}$. The results are presented in Fig.\ref{fig5}.
Since the BFO remains metallic within this approximation, we can
classify the main magnetic interaction as the double exchange. The
consequence of the strong ferromagnetic coupling is the largely
overestimated $T_C$ compared to the measured values. Although applying
the GGA+U method reduces slightly the exchange interaction, mainly due
to the reduction of the overlap between the Fe and oxygen states, the
Curie temperature is still too high in the whole range of
$U_{eff}$. The dependence of $T_C$ on $U_{eff}$ is nontrivial. First,
it reduces with the increase of $U_{eff}$, since the Fe $3d$ states
are getting localized and their overlap with $sp$ states of oxygen. At
$U_{eff}=2$~eV, a band gap forms in the minority spin channel and the
$J^{Fe-Fe}$ rise up with the increase of $U_{eff}$. Therewith, the
exchange interaction between the nearest neighbors is getting larger
only until $U_{eff}\approx 4$~eV and then decreasing, while the next
nearest neighbor interaction continues to increase until
$U_{eff}\approx 6$~eV. As result, the Curie temperature gets up to
700~K within the RPA for $U_{eff}\approx 6$~eV and then starts to fall
down. According to our estimations, the Curie temperature agrees with
experimental values only for $U_{eff}> 15$~eV, which are nonphysical
for this compound. Thus, consideration only electronic correlation
effects does not explain magnetic properties of BFO. This fact
motivated us to improve the model of BFO and make further
calculations.

\subsection{Simulation of XMCD spectra}
In complex transition metal oxides, the XAS and XMCD spectra at the
$L_{2,3}$ absorption edges can be used as fingerprints of the ground
state. In our study we simulated the XAS and XMCD spectra varying the
crystal structure, the chemical composition, the value of $U_{eff}$
for electronic correlation effects, the magnetic order,
and searching for optimal agreement with available experiments.

\begin{figure}[tbp!]
\begin{center}
  \includegraphics[width = .95\columnwidth]{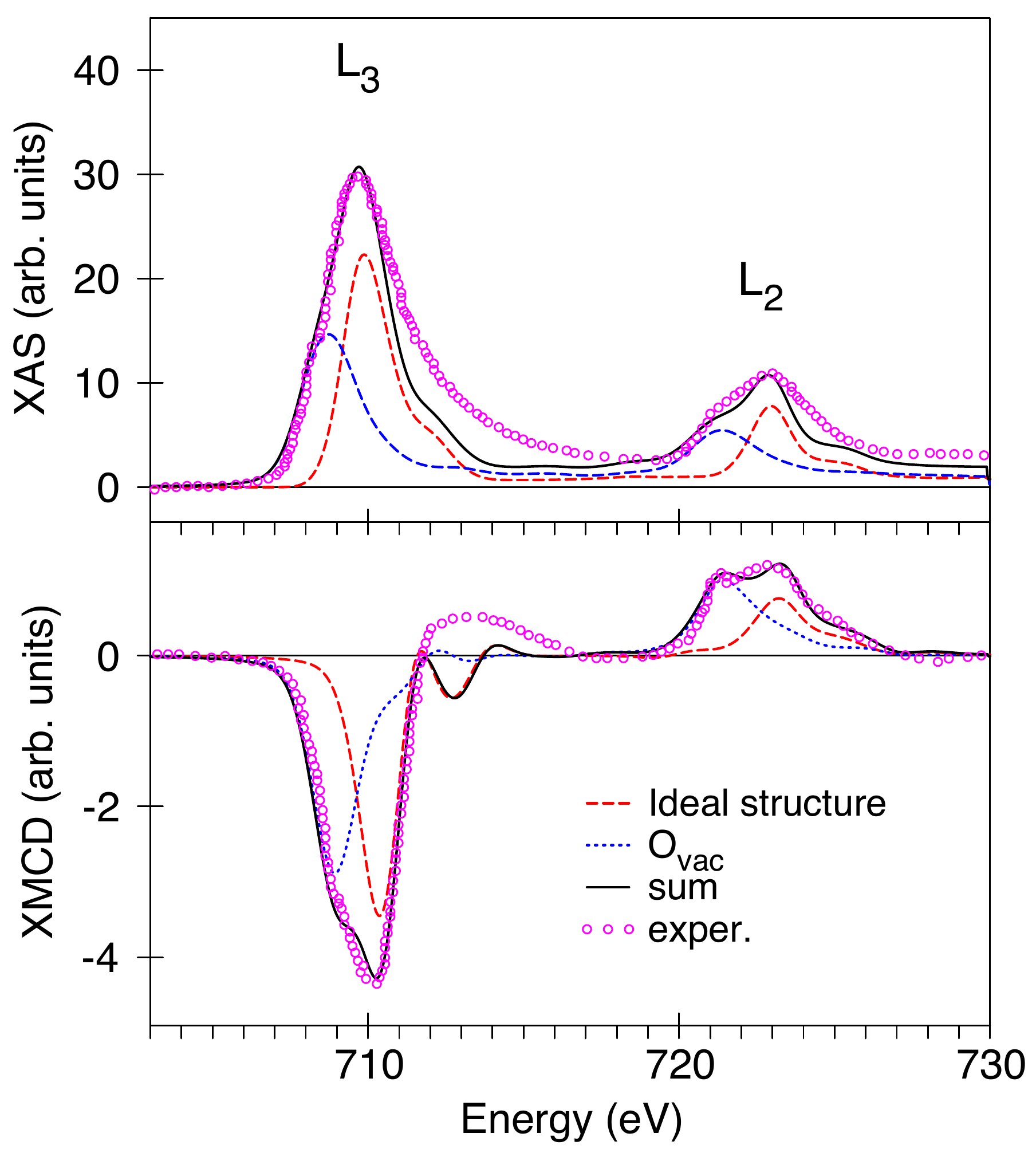}
\end{center}
\caption{\label{fig6} (Color) Comparison between the
  experimental \cite{TMC+15} x-ray absorption (upper panel) and x-ray
  magnetic circular dichroism (lower panel) at the Fe $L_{2,3}$ edges
  of BaFeO$_3$ and the spectra calculated in the GGA+U approximation
  ($U_{eff}$=3~eV) (black solid lines). Red dashed line show the
  spectra for ideal crystal structure, dotted blue lines present the
  theoretical spectra with oxygen vacancy. }
\end{figure}

Fig.~\ref{fig6} displays the experimentally measured \cite{TMC+15} and
theoretically calculated Fe $L_{2,3}$ XAS (upper panel) and XMCD
(lower panel) spectra of the cubic BaFeO$_3$. The $L_3$ XAS spectrum
exhibits a weak lower-energy shoulder at around 708~eV together with a
high-energy shoulder at 713~eV. The $L_2$ XAS spectrum exhibits a
lower-energy peak at around 721~eV. The Fe $L_3$ XMCD spectrum
consists of major negative peak at 710~eV with low-energy shoulder and
an additional positive peak at 713~eV. The Fe $L_2$ XMCD spectrum has
double-peak structure.

First, we found the best agreement with experiments was achieved for
$U_{eff}=3$~eV. This value of $U_{eff}$ is used later for all our
calculations if it is not mentioned specifically. However, the
experimental results could not explained only by taking into account
strong correlations effect. For the ideal cubic BFO structure, our
simulations of the XAS and XMCD spectra provide x-ray absorption
intensity in the Fe $L_3$ only at the major peak and high-energy
shoulder, while the low energy peaks at the $L_3$ and $L_2$ edges is not
reproduced (red dashed curve in Fig.~\ref{fig6}). The calculation
fails to reproduce low energy shoulder of the XMCD major negative peak
at 709~eV in $L_3$ XMCD spectrum. Also, the simulations for the ideal cubic BFO phase provide
only one a high energy peak structure in the $L_2$ XMCD spectrum,
however, the experimental measurements exhibit a double-peak
structure. 

The correct explanation of the experimental XAS and XMCD spectra is
only possible by taking into account crystal imperfections, namely,
oxygen vacancy. To investigate the influence of oxygen deficiency on
the XAS and XMCD spectra we created an oxygen vacancy in a double
supercell in the first neighborhood of the second Fe atom. We
found that the presence of the oxygen vacancy decreases the valency of
nearest Fe ion from 4+ to 3+ (see the corresponding discussion
below). The x-ray absorption from the Fe atoms with the oxygen
vacancy in close vicinity (dotted blue line in Fig.~\ref{fig6})
contributes to the low energy peaks in the $L_2$ absorption and XMCD
spectra. The oxygen vacancy also explain the existence of the
low-energy shoulder in Fe $L_3$ XMCD spectrum at 709~eV.

\begin{figure}[tbp!]
\begin{center}
  \includegraphics[width = .95\columnwidth]{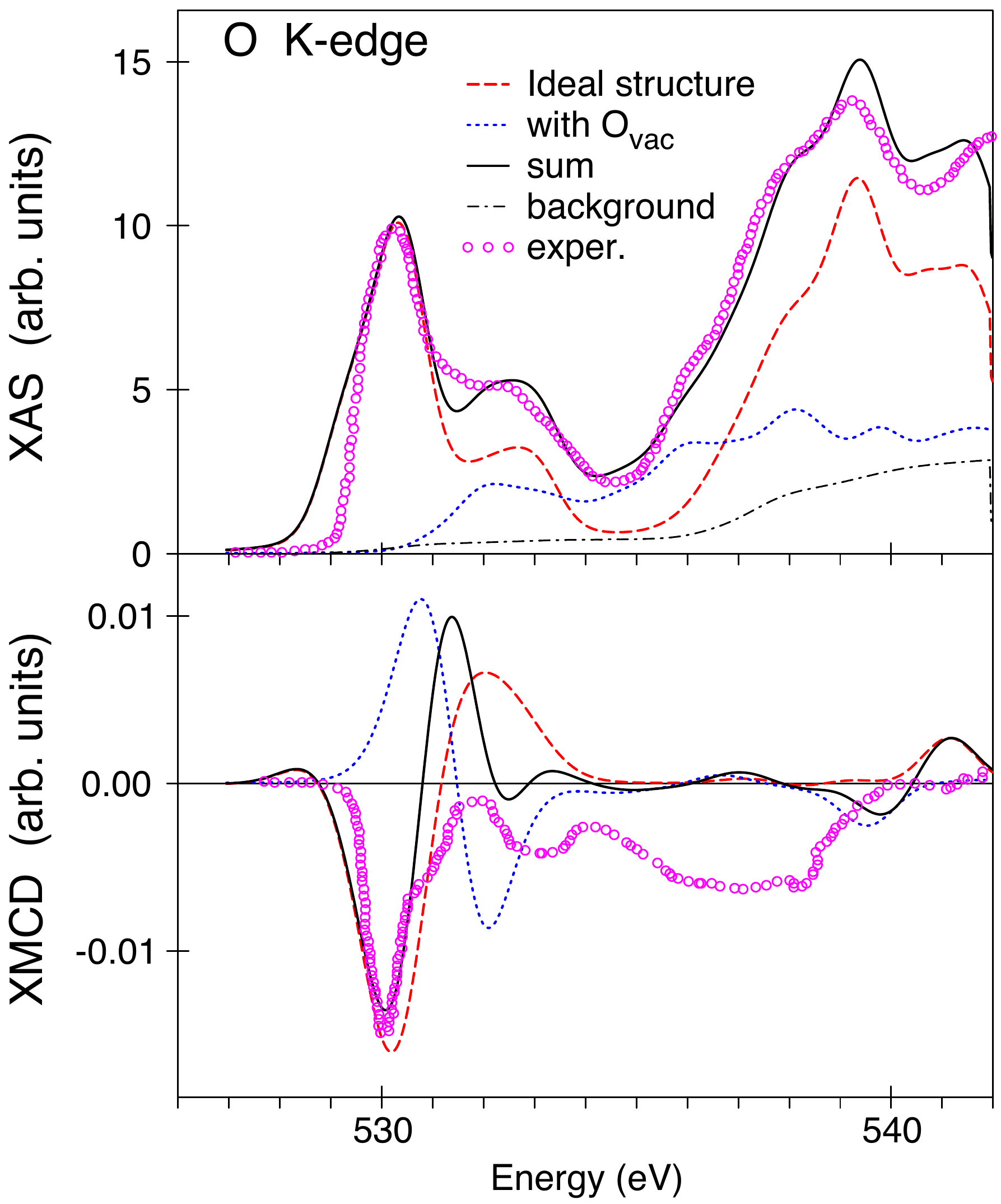}
\end{center}
\caption{\label{fig7} (Color) Comparison between the
  experimental \cite{TMC+15} (circles) x-ray absorption (upper panel)
  and x-ray magnetic circular dichroism (lower panel) at the O $K$
  edges of BaFeO$_{3}$ and the spectra calculated by the LMTO
  method in the GGA+U approximation ($U_{eff}$=3~eV) (black solid
  lines). Red dashed line show the spectra for ideal crystal
  structure, dotted blue lines present the theoretical spectra with
  oxygen vacancy. }
\end{figure}

The same conclusion can be drawn from the interpretation of the XAS and
XMCD spectra at the O $K$ edge.  Fig.\ref{fig7} shows the
experimentally measured x-ray absorption and XMCD spectra
\cite{TMC+15} (open circles) at the O $K$ edge in BaFeO$_3$ together with
theoretically calculated spectra for the ideal crystal structure of
BaFeO$_3$ (red dashed curves) and the structure with oxygen vacancy
(dotted blue curves). The calculations for the ideal structure are not
able to describe the high-energy shoulder of major negative XMCD peak
at 531~eV, which is due to the oxygen vacancy. We can conclude that
the explanation of the experimental spectra at both the Fe $L_{2,3}$
and O $K$ edges demands the presence of crystal imperfections in
BaFeO$_3$.

\subsection{Fe valency and oxygen vacancies in BaFeO$_{3-\delta}$}
To understand the vacancy formation we performed first-principles
study of the Fe oxidation states in the cubic BaFeO$_3$ using a
self-interaction method as it is described in Sec.~\ref{sec2}.  The Fe
oxidation state vary in different compounds from 2+ to 4+.  For a
completely oxidized BFO, one can expect a high valency, Fe$^{4+}$.
The total energy study of the oxidation state and further comparison
between BFO and BaFeO$_{3-\delta}$ can provide us with some hints
concerning the electronic properties and the preferable magnetic order
in BFO.  The Goodenough-Kanamori rules of the 180$^{\circ}$
superexchange suggest that the interaction Fe--O--Fe can be switched
from antiferromagnetic to ferromagnetic when the Fe 3$d$-orbital
filling changes. In this context, a variety of the superexchange
options, such as Fe$^{4+}$--O--Fe$^{4+}$, Fe$^{3+}$--O--Fe$^{3+}$ and
Fe$^{4+}$--O--Fe$^{3+}$ need to be inspected.  However, the
Goodenough-Kanamori simplifications work robustly for insulators only.
When theory deals with metallic magnetic oxide, such as the cubic BFO,
the superexchange coupling does not dominate there and double exchange
complicates the long-range interactions between magnetic species.
 
\begin{figure}
  \centering
  \includegraphics[width = .95\columnwidth]{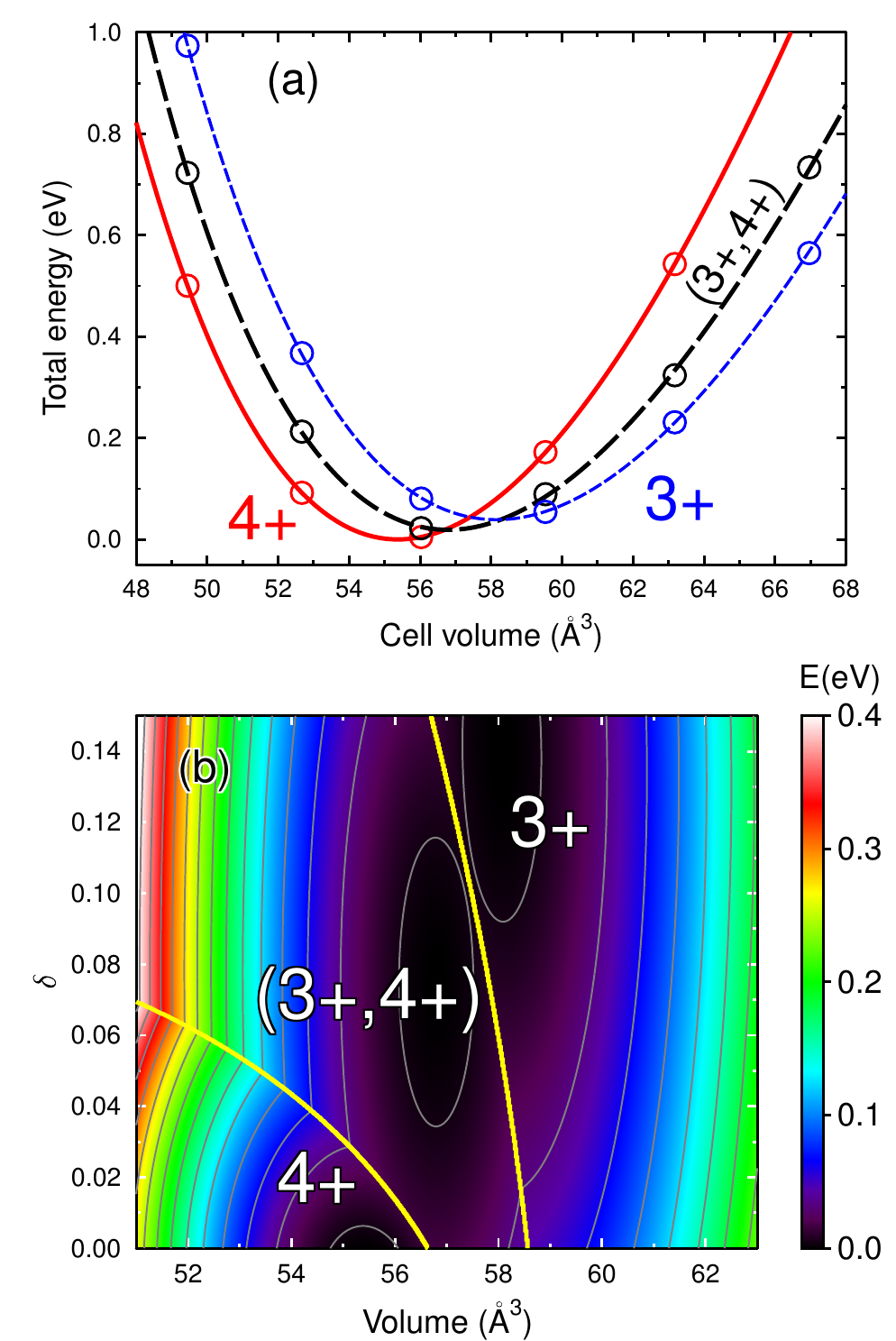}
  \caption{(Color) The total energies of BaFeO$_{3-\delta}$, calculated using the LSDA-SIC option for the Fe$^{4+}$, Fe$^{3+}$, and a combination of them 1:1. Energies as functions of the volume without oxygen
 vacancies (a).
Total energy as a function of volume and $\delta$ (b).}
  \label{fig8}
\end{figure}

In the case of the cubic symmetry, the Fe$^{4+}$ electronic
configuration is $t_{2g}^3 e_g^1$, where the half-occupied $e_g$
orbital has ether $d_{3z^2-r^2}$ or $d_{x^2-y^2}$ representation.  For
the cubic BFO case, which was treated within the LSDA-SIC model, we
found that the two $e_g$ options result in the same equilibrium volume
(see Fig.\ref{fig8}), which is substantially smaller then the
experimental result (62.57~\AA$^3$) and the volume calculated using the GGA
(Fe$^{2+}$). Although  the volume computed within the GGA is very
close to the experimental value, the corresponding total energy is
about 22~meV higher then the total energies of  Fe$^{4+}$ and  Fe$^{3+}$
configurations. 

For the cation Fe$^{3+}$, its t$_{2g}$ and e$_g$ levels must be fully
occupied in the majority spin channel.  This electronic configuration
yields the total-energy minimum value, which is not far above that of
the Fe$^{4+}$ ground state, as shown in Fig.~\ref{fig8}. The
calculated equilibrium volume is in a better agreement with the
experimental results then the values obtained for Fe$^{3+}$ (SIC) and
Fe$^{2+}$ (GGA) configurations. The energetics suggest that both
oxidations, Fe$^{4+}$ and Fe$^{3+}$, can be realized in epitaxially
fabricated thin films. Configurations with other oxidation states are
much higher in energy and are excluded from the further consideration.

Further, we compared the total energies of Fe$^{4+}$ and Fe$^{3+}$
oxidation states in the presence of oxygen vacancies. For
$\delta=0.03$ the total energies in the equilibrium volumes are almost
the same (see Fig.~\ref{fig8}). Thus, the vacancy formation lowers the
energy of the Fe$^{3+}$ oxidation state, which agrees with our
interpretation of the XAS and XMCD experiments. In a real BFO sample,
both oxidation states can coexist, which is as well evident from our
XAS and XMCD simulations. In Fig.~\ref{fig8} we present the total
energy for a mixed configuration, which is lower then the others in a
particular concentration range.

\begin{figure}
  \centering
  \includegraphics[width = .6\columnwidth]{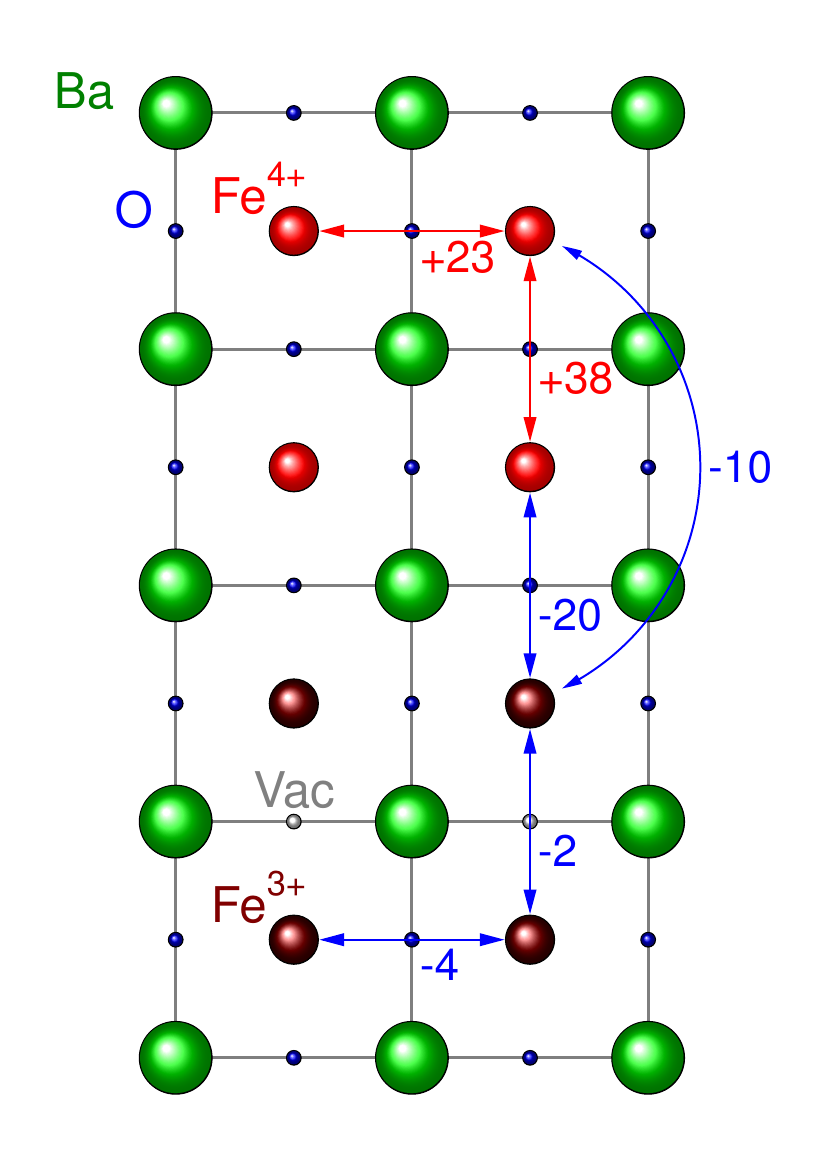}
  \caption{(Color) Exchange interaction in the BaFeO$_{2.75}$
    simulated in a $4\times 1 \times 1$ supercell.}
  \label{fig9}
\end{figure} 

\begin{figure}
  \centering
  \includegraphics[width = .95\columnwidth]{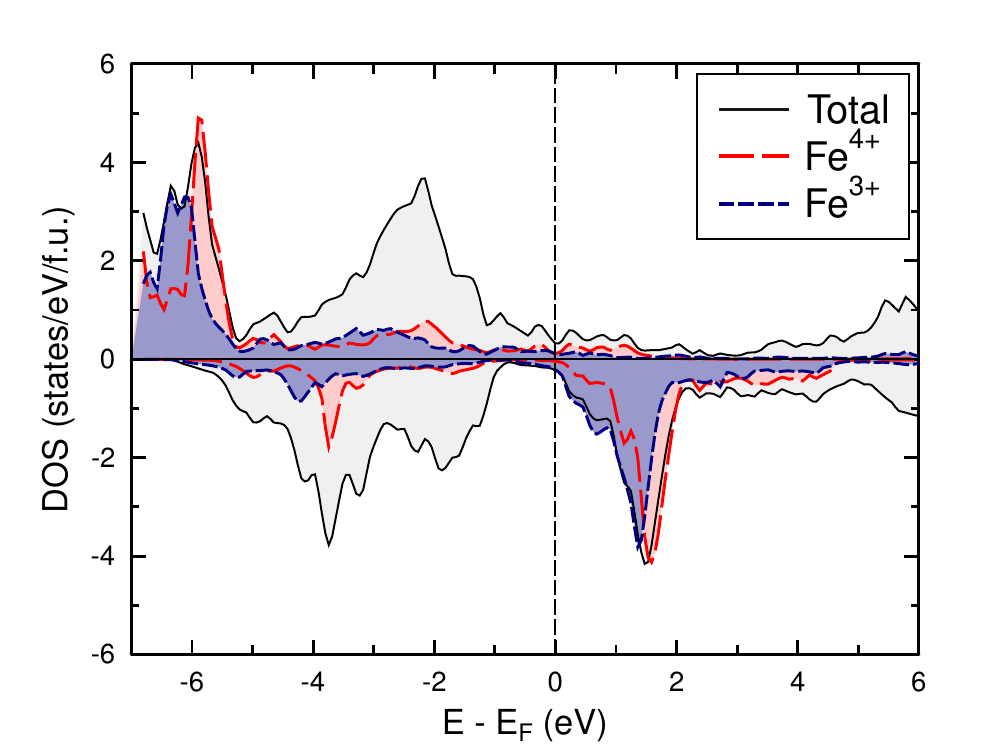}
  \caption{(Color) DOS of the BaFeO$_{2.75}$
    simulated in a $4\times 1 \times 1$ supercell.}
  \label{fig10}
\end{figure} 

As next, we investigated the impact of oxygen vacancies on magnetic
properties of the BFO.  The oxygen vacancies were simulated within a
$4\times 1 \times 1$ supercell, in which one oxygen site was occupied by an
vacancy with a given probability. The disorder was simulated using the
CPA method.  The use of the supercell enables to take into account
short-range effects in a simple manner, which can take place in real
materials. Results for the exchange interaction, computed in this
supercell and $\delta=0.25$, are presented in Fig.~\ref{fig9}. While
the exchange interaction between the Fe$^{4+}$ cations is strong and
positive, it is week and negative between the Fe$^{3+}$ atoms. The
reason for this can be understood from the DOS, shown in
Fig.~\ref{fig10}. Because of the vacancy (missing oxygen), the $3d$
states of the Fe$^{3+}$ are more localized (in particular the $e_{g}$
states) and are shifted down in energy. This increases the tendency to
an antiferromagnetic coupling. The coupling between the atoms in
different oxidation states is as well antiferromagnetic and very
strong due to the large difference of the DOS.
This funding agrees with
the main conclusions of S. Mori, who first developed a model for
the magnetic structure of BaFeO$_x$.\cite{Mori1970}

\begin{figure}
  \centering
  \includegraphics[width = .95\columnwidth]{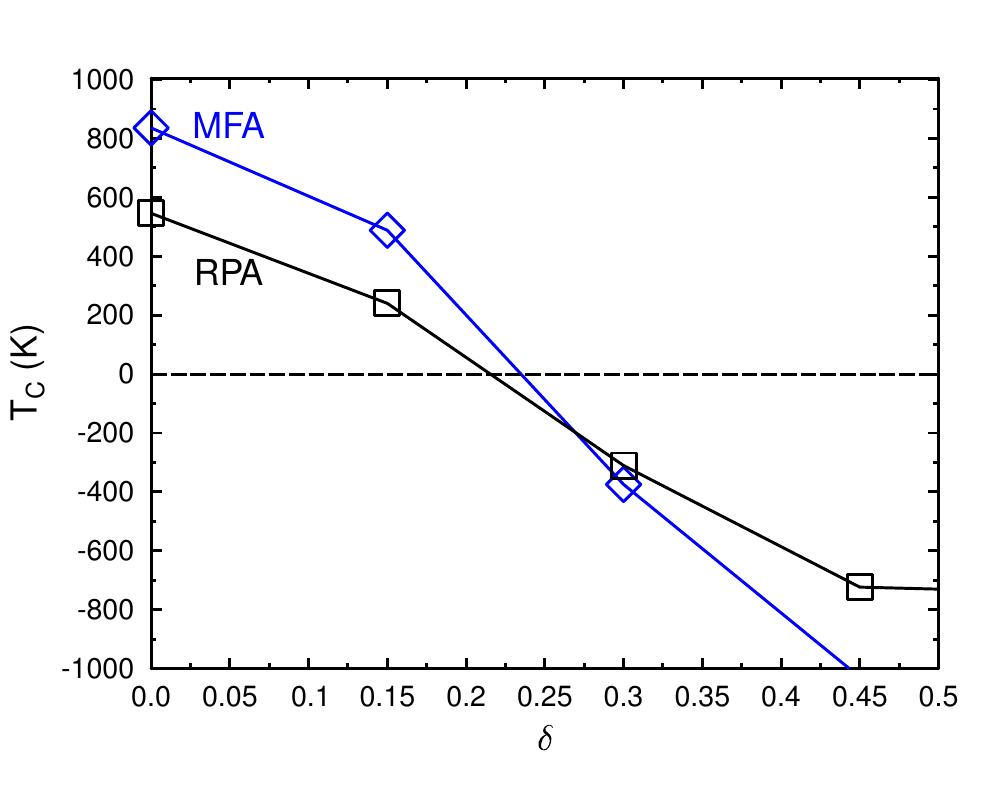}
  \caption{(Color) Critical temperature $T_C$ in BaFeO$_{3-\delta}$
    as a function of oxygen vacancy concentration $\delta$. Here, the
    negative $T_C$ means the absence of the ferromagnetic order.}
  \label{fig11}
\end{figure}

Depending on the vacancy concentration and their distribution, the
non-trivial exchange interaction in BaFeO$_{3-\delta}$ can results in
various magnetic orders. In Fig.~\ref{fig11}, we show the change of
critical temperature with the oxygen vacancy concentration simulated
within the MFA and RPA approaches. The negative $T_C$ means here the
absence of the ferromagnetic order (either antiferromagnetic or
non-collinear). Thus, for a moderate vacancy concentration $\delta$ of
0.15 -- 0.20, $T_C$ lies in the experimental range of 250 -- 100~K.

\section{Summary}
In this work we studied electronic and magnetic properties of a cubic
BaFeO$_3$ within a DFT framework using a GGA and a GGA+U approaches. We
predicted that the BaFeO$_3$ in an ideal stoichiometric cubic structure
should be half-metallic and ferromagnetic with a high Curie
temperature of 700 -- 900~K. However, this finding disagrees the
experiments, which found that the compound has a $T_C$ far below the room
temperature. To
understand this discrepancy we simulated XAS and XMCD spectra
comparing the theoretically obtained results with available
experimental data. Herewith, we varied the chemical composition,
structural and electronic parameters to obtain a better agreement with
experiments. Our simulations showed that the experimental observations
can be explained by oxygen vacancies, which can be present in
real BFO samples. Thereby, Fe atoms near an oxygen vacancy change
their oxidation state from $4+$ to $3+$. Thus, a real cubic BaFeO$_3$
may contain both Fe$^{4+}$ and Fe$^{3+}$ cations, which explains the
experimentally observed double peak structure in XAS and XMCD
spectra. Further, we found that oxygen vacancies can substantially
reduce the Curie temperature of a cubic BFO or even change the
ferromagnetic order to an antiferromagnetic one or a non-collinear
structure. Responsible for this is a strong antiferromagnetic exchange
coupling between Fe atoms of different oxidation states. This result
provides a clear route to fabricate a robust ferromagnetic BaFeO$_3$
with a high Curie temperature by reducing the amount of oxygen
vacancies and stabilizing it in the cubic structure. 

\section{Acknowledgments}

 This work is supported by the DFG within the Collaborative Research Center 
\textit{Sonderforschungsbereich} SFB 762, 'Functionality of Oxide Interfaces'.

\end{document}